\documentstyle[12pt,epsfig]{article}
\begin{document}

\begin{titlepage}
\rightline{\large October 2005}
\vskip 2cm
\centerline{\Large \bf Implications of the DAMA/NaI and CDMS experiments 
}
\vskip 0.2cm
\centerline{\Large \bf for mirror matter-type dark matter}
\vskip 2.2cm
\centerline{\large R. Foot\footnote{
E-mail address: rfoot@unimelb.edu.au}}

\vskip 0.7cm
\centerline{\large \it School of Physics,}
\centerline{\large \it University of Melbourne,}
\centerline{\large \it Victoria 3010 Australia}
\vskip 2cm
\noindent
We re-analyse the implications of the 
DAMA/NaI experiment for
mirror matter-type dark matter,
taking into account information from the energy dependence
of the DAMA annual modulation signal. This is combined with the null
results from the CDMS experiment, leading to fairly
well defined allowed regions of parameter space.
The allowed regions of parameter space will be probed in the near
future by the DAMA/LIBRA, CDMS, and other experiments,
which should either exclude or confirm this explanation
of the DAMA/NaI annual modulation signal.
In particular, we predict that the CDMS experiments should find
a positive signal around the threshold recoil energy region,
$E_R < 15 \ {\rm keV}$ in the near future.

\end{titlepage}

\section{Introduction}

The exact parity symmetric model\cite{flv} is the minimal extension of
the standard model which allows for an exact unbroken parity symmetry
[$x \to -x$, $t \to t$].
According to this theory, each type of ordinary
particle (electron, quark, photon etc) has a corresponding mirror partner
(mirror electron, mirror quark, mirror photon etc), 
of the same mass. The two sets of particles form 
parallel sectors each with gauge symmetry $G$
(where $G = SU(3) \otimes SU(2) \otimes U(1)$ in the 
simplest case)
so that the full gauge group is $G \otimes G$.
The unbroken mirror symmetry maps
$x \to -x$ as well as ordinary particles into mirror
particles. Exact unbroken time reversal symmetry
also exists, with standard CPT identified as the product
of exact T and exact P\cite{flv}.

It has been argued that the stable mirror particles, mirror nucleons and
mirror electrons are an interesting candidate for the inferred
dark matter of the Universe (for a review, see Ref.\cite{review}).
Of course, to be a successful dark matter candidate, mirror
matter needs to behave, macroscopically, differently to ordinary matter.
In particular, four key distinctions need to be explained:
\begin{itemize}
\item
The cosmological abundance of mirror matter should be different to
ordinary matter,
$\Omega_{dark} \neq \Omega_{matter}$. 
\item
Mirror particles
should give negligible contribution to the energy density at the
epoch of big bang nucleosynthesis.
\item
Structure formation in the mirror
sector must begin before ordinary matter radiation decoupling.
\item
In spiral galaxies,
the time scale for the collapse of ordinary matter onto the disk
must be much shorter
than that of mirror matter.

\end{itemize}

Clearly, mirror matter behaves differently to ordinary matter,
at least macroscopically.
It is hypothesised that
this macroscopic  asymmetry originates from
effectively
different initial conditions in the two sectors. The exact
mircopscopic (Lagrangian) symmetry between ordinary and mirror
matter need never be broken. In particular, if ordinary
and mirror particles have different temperatures in the
early Universe, $T' \ll T$, then
the mirror particles give negligible contribution to the energy
density at the time of nucleosynthesis leading to standard big
bang nucleosynthesis. Another consequece of $T' \ll T$ is that
mirror photon decoupling occurs earlier than ordinary photon 
decoupling -- implying that mirror structure formation can begin
before ordinary photon decoupling. 
In this way, mirror matter-type dark matter can successfully explain
the large scale structure formation (for detailed studies, 
see ref.\cite{comelli,lss,other}).  
Also, $\Omega_{dark} \neq
\Omega_{matter}$ could also be due to different effective initial conditions
in the early Universe (see e.g. ref.\cite{new1},\cite{new2} for some
specific scenarios).

If mirror matter is the inferred non-baryonic dark matter
in the Universe, then
the halo of our galaxy
should be gas of ionized mirror atoms and mirror electrons
together with a non-gaseous component of $f \sim 0.2$ (which
can be inferred from gravitational microlensing studies\cite{macho}).
Although dissipative, roughly spherical galactic mirror matter halo's can exist without
collapsing provided that a heating mechanism exists - with
ordinary and/or supernova explosions being plausible
candidates\cite{sph}. 
Obviously, the heating of the ordinary and mirror matter
in spiral galaxies needs to be asymmetric, but again, due to
different initial conditions in the early Universe, asymmetric heating
is plausible. For example, the early Universe temperature asymmetry, 
$T' \ll T$ (expected from successful big bang nucleosynthesis and Large
scale structure formation, as discussed above) 
implies that the primordial mirror
helium/mirror hydrogen ratio will be much larger than the corresponding
ordinary helium/ordinary hydrogen ratio\cite{comelli}. 
Consequently the formation and evolution of stars in the ordinary
and mirror sectors are completely different.
The details of
the evolution on (sub) galactic scales, is of course, very complex, and is  
yet to be fully understood. 

Ordinary and mirror particles interact with each other
by gravity and via photon-mirror photon kinetic
mixing:\footnote{
Technically, photon-mirror photon kinetic mixing arises from
kinetic mixing of the $U(1)$ and $U(1)'$ gauge fields, since
only for the abelian $U(1)$ gauge symmetry is such mixing
gauge invariant\cite{fh}.
The only other gauge invariant and renormalizable interactions mixing
ordinary and mirror particles are the quartic Higgs - mirror Higgs interaction:
$\lambda \phi^{\dagger}\phi \phi'^{\dagger}\phi'$ and neutrino - mirror
neutrino mass mixing\cite{flv,flv2}.}
\begin{eqnarray}
{\cal L} = {\epsilon \over 2}
F^{\mu \nu} F'_{\mu \nu}
\end{eqnarray}
where $F^{\mu \nu}$ ($F'_{\mu \nu}$) is the field
strength tensor for electromagnetism (mirror electromagnetism).
One effect of photon-mirror photon kinetic mixing is to cause mirror
charged particles (such as the mirror proton and mirror electron)
to couple to ordinary photons with effective electric charge $\epsilon e$.\cite{flv,holdom,sasha}
The various experimental implications of photon-mirror photon kinetic 
mixing have been
reviewed in Ref.\cite{freview}.
Of most relevance for this paper, is that
this interaction enables mirror particles 
to elastically scatter off ordinary particles -- essentially
Rutherford scattering.

A detector on Earth can therefore be used to detect halo mirror
nuclei via elastic scattering. Several previous
papers\cite{f1,f2,f3} have explored this possibility, especially in view
of the impressive dark matter signal from the DAMA/NaI experiment\cite{dama}. 
The purpose of this paper is to re-analyse the mirror matter interpretation of
the DAMA/NaI experiment in combination with the null results of the CDMS
experiment\cite{cdms} in order to pin down more precisely the currently favoured
region of parameter space within this scenario.

The outline of this paper is as follows:
In section 2
we review and extend 
the Mirror matter interpretation of the positive DAMA/NaI annual 
modulation signal.
In section 3 we utilize the measured
recoil energy dependence of the DAMA/NaI annual modulation signal
to constrain the mirror matter interpretation. In section 4 we
then examine the implications of the null results obtained in the
CDMS/Ge
experiment. Importantly, we  
find that there is a fairly significant allowed region 
of parameter space consistent with the positive DAMA/NaI annual
modulation signal and the null results of the CDMS experiment.
Finally in section 5 we conclude.

\section{Mirror dark matter implications for direct detection 
experiments such as DAMA/NaI}

Let us first briefly review the required technology (see references\cite{f1,f2,f3} for
more details).
For definiteness, consider
a halo mirror nuclei, $A'$, of atomic number $Z'$
scattering off an ordinary nucleus, $A$ (in an ordinary matter detector) 
of atomic number $Z$.
The cross section is then just of the standard Rutherford form
corresponding to a particle of electric charge $Ze$ scattering
with a particle of electric charge $\epsilon Z' e$.
This cross section can be expressed in terms of the recoil
energy of the ordinary atom, $E_R$, and 
the velocity in the Earth's rest frame, $v$:
\begin{eqnarray}
{d\sigma \over dE_R} = {\lambda \over E_R^2 v^2}
\label{cs}
\end{eqnarray}
where 
\begin{eqnarray}
\lambda \equiv {2\pi \epsilon^2 \alpha^2 Z^2 Z'^2 \over M_A} \
F_{A}^2 (qr_A) F_{A'}^2 (qr_{A'})
\end{eqnarray}
and $F_X (qr_X)$ ($X = A, A'$) are the form factors which 
take into account the finite size of the nuclei and mirror nuclei.
[$q = (2M_A E_R)^{1/2}$ is the momentum transfer and $r_X$ is the effective
nuclear radius]\footnote{
We use natural units, $\hbar = c = 1$ throughout.}.
A simple analytic expression for
the form factor, which we adopt in our numerical work, is the one
given by Helm\cite{helm,smith}:
\begin{eqnarray}
F_X (qr_X) = 3{j_1 (qr_X) \over qr_X} e^{-(qs)^2/2}
\end{eqnarray}
with $r_X = 1.14 X^{1/3}$ fm, $s = 0.9$ fm.
In this equation, $j_1$ is the spherical Bessel function of index 1.

In an experiment such as DAMA/NaI\cite{dama}, the measured quantity is 
the recoil energy, $E_R$, of a target atom. 
The interaction rate is
\begin{eqnarray}
{dR \over dE_R} &=& 
\sum_{A'} N_T n_{A'} \int {d\sigma \over dE_R} {f_{A'}(v,v_E) \over k} |v|
d^3v \nonumber \\
&=& \sum_{A'} N_T n_{A'}
{\lambda \over E_R^2 } \int^{\infty}_{|v| > v_{min}
(E_R)} {f_{A'}(v,v_E) \over k|v|} d^3 v 
\label{55}
\end{eqnarray}
where $N_T$
is the number of target atoms per kg of detector\footnote{
For detectors with more than one target element 
we must work out the
event rate for each element separately and add them up to get the total
event rate.}. Also,
$n_{A'}$ is the halo number density (at the Earth's location) 
of the mirror element, $A'$ and 
$f_{A'}(v,v_E)/k$ is its velocity distribution 
($k$ is the normalization factor)
with $v$ being the velocity relative
to the Earth, and $v_E$ is the Earth velocity relative to the
dark matter distribution. The lower velocity limit,
$v_{min} (E_R)$, 
is given by the kinematic relation:
\begin{eqnarray}
v_{min} &=& \sqrt{ {(M_A + M_{A'})^2 E_R\over 2M_A M^2_{A'}} } .
\label{v}
\end{eqnarray}

Considering a particular mirror chemical element, $A'$ (e.g.
$A' = H', He', O'$ etc), the velocity distribution for
these halo mirror particles is then:
\begin{eqnarray}
f_{A'} (v, v_E) = exp\left[ -{1 \over 2} M_{A'} (v+v_E)^2/T \right]
= exp[-(v+v_E)^2/v_0^2]
\label{d12}
\end{eqnarray}
where $v_0^2 \equiv 2T/M_{A'}$. The assumption of approximate
hydrostatic equilibrium for the halo particles implies
a relation between $T$ and the local rotational velocity, 
$v_{rot}$:\cite{f2}
\begin{eqnarray}
T = {\mu M_p v_{rot}^2 \over 2}
\label{d13}
\end{eqnarray}
where $\mu M_p$ is the mean mass of the particles
comprising the mirror (gas) component of the halo ($M_p$ is the
proton mass).
Note that the Maxwellian distribution should be an excellent
approximation in the case of mirror dark matter, since
the self interactions of the particles ensure that halo is
thermalized.

The velocity integral in Eq.(\ref{55}),
\begin{eqnarray}
I (E_R) \equiv \int^{\infty}_{|v| > v_{min}(E_R)} {f_{A'}(v,v_E) \over k|v|} d^3 v
\end{eqnarray}
is standard (similar integrals occur in the usual
WIMP interpretation\footnote{
However in the WIMP case the upper velocity limit is finite, 
corresponding to the galactic escape velocity. While for
mirror dark matter, the upper limit is infinite due
to the self interactions of the mirror particles.} 
) and can easily be evaluated in terms of
error functions assuming
a Maxwellian dark matter distribution\cite{smith},
$f_{A'}(v,v_E)/k = (\pi v_0^2)^{-3/2} \ exp[-(v+v_E)^2/v_0^2]$,
\begin{eqnarray}
I(E_R) = {1 \over 2v_0 y}\left[ erf(x+y) - erf(x-y)\right] 
\label{ier}
\end{eqnarray}
where 
\begin{eqnarray}
x \equiv {v_{min} (E_R) \over v_0}, \ y \equiv {v_E \over v_0}.
\label{xy}
\end{eqnarray}
The Earth's velocity relative to the galaxy, $v_E$, has
an estimated mean value  of
$\langle v_E \rangle \simeq v_{rot} + 12 \ {\rm km/s}$, 
with  $v_{rot}$, the local
rotational velocity, in the
$90\%$ C.L. range\cite{koch},
\begin{eqnarray}
170 \ {\rm km/s} \stackrel{<}{\sim} v_{rot} \stackrel{<}{\sim} 270\ {\rm km/s}.
\label{range2}
\end{eqnarray}
While some estimates put more narrow limits on the local
rotational velocity, it is useful to allow for
a broad range for $v_{rot}$ since it can also
approximate the effect of bulk halo rotation.

As can be seen from Eq.(\ref{d12},\ref{d13}),
in the case of mirror matter-type dark matter,
the $v_0$ value
for a particular halo component element, $A'$,
depends on the chemical composition of the halo.
In general,
\begin{eqnarray}
{v_0^2 (A') \over v_{rot}^2} = {\mu M_p \over M_{A'}}
\label{z3}
\end{eqnarray}
The most abundant mirror elements
are expected to be $H', He'$, generated in
the early Universe from mirror big bang nucleosynthesis
(heavier mirror elements should be generated in mirror
stars). It is useful, therefore, to consider
two limiting cases: first that the halo
is dominated by $He'$ and the second is that
the halo is dominated by $H'$. The mean mass of
the particles in the halo are then
(taking into
account that the light halo mirror atoms should be fully ionized):
\begin{eqnarray}
\mu M_p & \simeq & M_{He'}/3 \simeq 1.3\ {\rm GeV \ for \ He' \ dominated \ halo,}
\nonumber \\
\mu M_p & \simeq  & M_{H'}/2 \simeq 0.5\ {\rm GeV \ for \ H' \ dominated \ halo.}
\end{eqnarray}
The $v_0$ values can then easily be obtained from Eq.(\ref{z3}):
\begin{eqnarray}
v_0 (A') &=& v_0 (He') \sqrt{{M_{He'}\over M_{A'}}} \approx {v_{rot} \over
\sqrt{3}}
\sqrt{{M_{He'}\over M_{A'}}} \ {\rm km/s} \ \  \ \rm{for \ He' \ dominated \
halo}\nonumber \\
v_0 (A') &=& v_0 (H') \sqrt{{M_{H'}\over M_{A'}}} \approx {v_{rot}
\over \sqrt{2}}
\sqrt{{M_{H'}\over M_{A'}}} \ {\rm km/s} \ \ \ \rm{for \ H' \ dominated \
halo}.
\end{eqnarray}
Mirror BBN\cite{comelli}
suggests that $He'$ dominates over $H'$, and this is what we assume
in our numerical work in this paper. However, it turns out
that the thresholds of the DAMA/NaI and CDMS experiments
are sufficiently high that these experiments are only sensitive to
mirror elements heavier than about carbon, which 
means that
$v_0 (A') \ll v_{rot}$ for these elements -- independently of 
whether $He'$ or $H'$ dominates the halo.
For this reason, our main results 
(such as the allowed regions in figure 4)
do not depend very significantly
on whether we assume that $He'$ or $H'$ dominates the mass
of the Halo.

The DAMA/NaI experiment\cite{dama} turns out to be very sensitive to mirror
matter-type dark matter because of the light target element, $Na$,
and the relatively low energy threshold of 2 keVee\footnote{
The unit, keVee is the so-called electron equivalent energy,
which is the energy of the event if due to an electron
recoil. The actual nuclear recoil energy
(in keV) is given by: keVee/$q$, where $q$ is the quenching factor
($q_I \simeq 0.09$ and $q_{Na} \simeq 0.30$).}.
This experiment uses the annual modulation signature\cite{sig},
which arises because of the Earth's motion around the sun.
The point is that the interaction rate, Eq.(\ref{55}), depends on $v_E$,
which varies due to the Earth's motion around the sun:
\begin{eqnarray}
v_E (t) &=& v_{\odot} + v_{\oplus} \cos\gamma \cos \omega (t - t_0)
\nonumber \\
&=& v_{\odot} + \Delta v_E \cos \omega (t - t_0)
\end{eqnarray}
where $v_{\odot} = v_{rot} + 12 \ {\rm km/s} \sim 230 \ {\rm km/s}$ is the sun's velocity
with respect to the galaxy and $v_{\oplus} \simeq 30$ km/s is the Earth's orbital
velocity around the Sun ($t_0 = 152.5$ days and 
$\omega = 2\pi/T$, with $T = 1$ year).
The inclination of the Earth's orbital plane relative
to the galactic plane is $\gamma \simeq 60^o$, which implies that 
$\Delta v_E \simeq 15$ km/s.
The differential interaction rate, Eq.(\ref{55}), 
can be expanded in a Taylor series around $v_E = v_{\odot}$, leading
to an annual modulation term:
\begin{eqnarray}
R_i = R^0_i + R^1_i \cos \omega (t - t_0)
\end{eqnarray}
where
\begin{eqnarray}
R^0_i &=& {1 \over \Delta E} \int^{E_i + \Delta E}_{E_i} \left( {dR \over dE_R}\right)_{v_E = v_{\odot}} \ dE_R,
\nonumber \\
R^1_i &=& {1 \over \Delta E} \int^{E_i + \Delta E}_{E_i} {\partial \over \partial v_E} \left(
{dR \over dE_R}\right)_{v_E = v_{\odot}} \ \Delta v_E dE_R
\end{eqnarray}
According to the DAMA analysis\cite{dama}, they indeed find an annual 
modulation at more than $6\sigma$ C.L.
Their data fit gives $T = (1.00 \pm 0.01)$ years and $t_0 = 140 \pm 22$ days, 
consistent
with the expected values. [The expected value for $t_0$ is 152.5 days (2 June), where
the Earth's velocity reaches a maximum with respect to the galaxy].
Their signal occurs in the 2-6 keVee energy range, with amplitude
\begin{eqnarray}
R^1 = (0.019 \pm 0.003) \ {\rm cpd/kg/keVee} \  {\rm [cpd = counts \ per \ day]}.
\label{kj}
\end{eqnarray}

The DAMA experiment itself is not sensitive to the dominant, 
$He'$ or $H'$ component.
These nuclei are too light to give a signal above the DAMA/NaI energy threshold.
DAMA is sensitive to mirror nuclei heavier than about carbon.
In this paper, we propose to approximate the
spectrum of such heavy mirror metals by
three components, $O', Si', Fe'$, which span the expected mass range.
In principle it might be possible to predict the relative
abundances of the mirror metal components if enough is known about the
initial conditions and stellar evolution in the mirror sector (for some 
preliminary work in this direction, see ref.\cite{recent}).
However, for the purposes of this paper, we leave the relative abundances of these
three components as free parameters to be fixed by the direct
detection experiments.

Interpreting the DAMA annual modulation signal [Eq.(\ref{kj})] in terms
of these three elements
and assuming a $He'$ dominated halo, we find numerically
that\cite{recent2}:

\begin{eqnarray}
|\epsilon | \sqrt{ {\xi_{O'} \over 0.10} +
{\xi_{Si'} \over 0.016} + {\xi_{Fe'} \over 0.015}}
\simeq 5.3^{+1.1}_{-1.4} \times 10^{-9}
\ {\rm for} \ v_{rot} = 220\ {\rm km/s},
\label{dama55}
\end{eqnarray}
where the errors denote a 3 sigma allowed range [i.e. $R^1 = 0.019 \pm 0.009$ cpd/kg/keVee] and
$\xi_{A'} \equiv n_{A'}M_{A'}/(0.3 \ {\rm GeV/cm^3})$
is the $A'$ proportion (by mass) of the halo dark matter.
Allowing for a range of $v_{rot}$, we find
\begin{eqnarray}
|\epsilon | \sqrt{ {\xi_{O'} \over 0.10} +
{\xi_{Si'} \over 0.004} + 
{\xi_{Fe'} \over 7.2\times 10^{-4}}}
\simeq 1.4^{+0.3}_{-0.4} \times 10^{-8}
\ \ {\rm for} \ v_{rot} = 170\ {\rm km/s}
\nonumber \\
|\epsilon | \sqrt{ {\xi_{O'} \over 0.10} +
{\xi_{Si'} \over 0.034} 
+ {\xi_{Fe'} \over 0.083}}
\simeq 4.3^{+0.9}_{-1.2} \times 10^{-9}
\ \ {\rm for} \ v_{rot} = 250\ {\rm km/s}
\label{dama65}
\end{eqnarray}
Evidently, the magnitude of $|\epsilon|\sqrt{\sum \xi_{A'}}$
depends somewhat on $v_{rot}$.

Note that mirror matter with $\epsilon \sim 10^{-9}$ 
has many interesting applications (see
e.g. ref. \cite{saibal,zurab}).
It is also consistent with Laboratory\cite{shelly} and
big bang nucleosynthesis constraints\cite{shelly2}.

\section{Recoil energy dependence of the annual modulation
signal}

The relative abundance of the $O', \ Si'$ and $Fe'$ components
can in principle be determined from
the annual modulation energy spectrum [ 
defined as $\frac{limit}{\Delta E \to 0} R^1 \equiv 
{\partial  \over \partial v_E}\left( {\partial R \over \partial
E}\right) \Delta v_E $]. 
In 
{\bf Figure 1a,b,c}, we give the predicted DAMA/NaI annual modulation energy
spectrum 
for $He'$ dominated halo 
assuming a mirror metal component consisting of 
a) pure $O'$ ($\xi_{Fe'} = \xi_{Si'} = 0$), b) pure $Si'$ ($\xi_{O'} = \xi_{Fe'} = 0$) 
and c) pure $Fe'$ ($\xi_{O'} = \xi_{Si'} = 0$), for
three representative values for $v_{rot}$.
In each case, $|\epsilon | \sqrt{\xi}$ is fixed so that $R^1 = 0.019$ cpd/kg/keVee in the $(2-6) \ 
{\rm keVee}$ region.

\vskip 0.3cm
\centerline{\epsfig{file=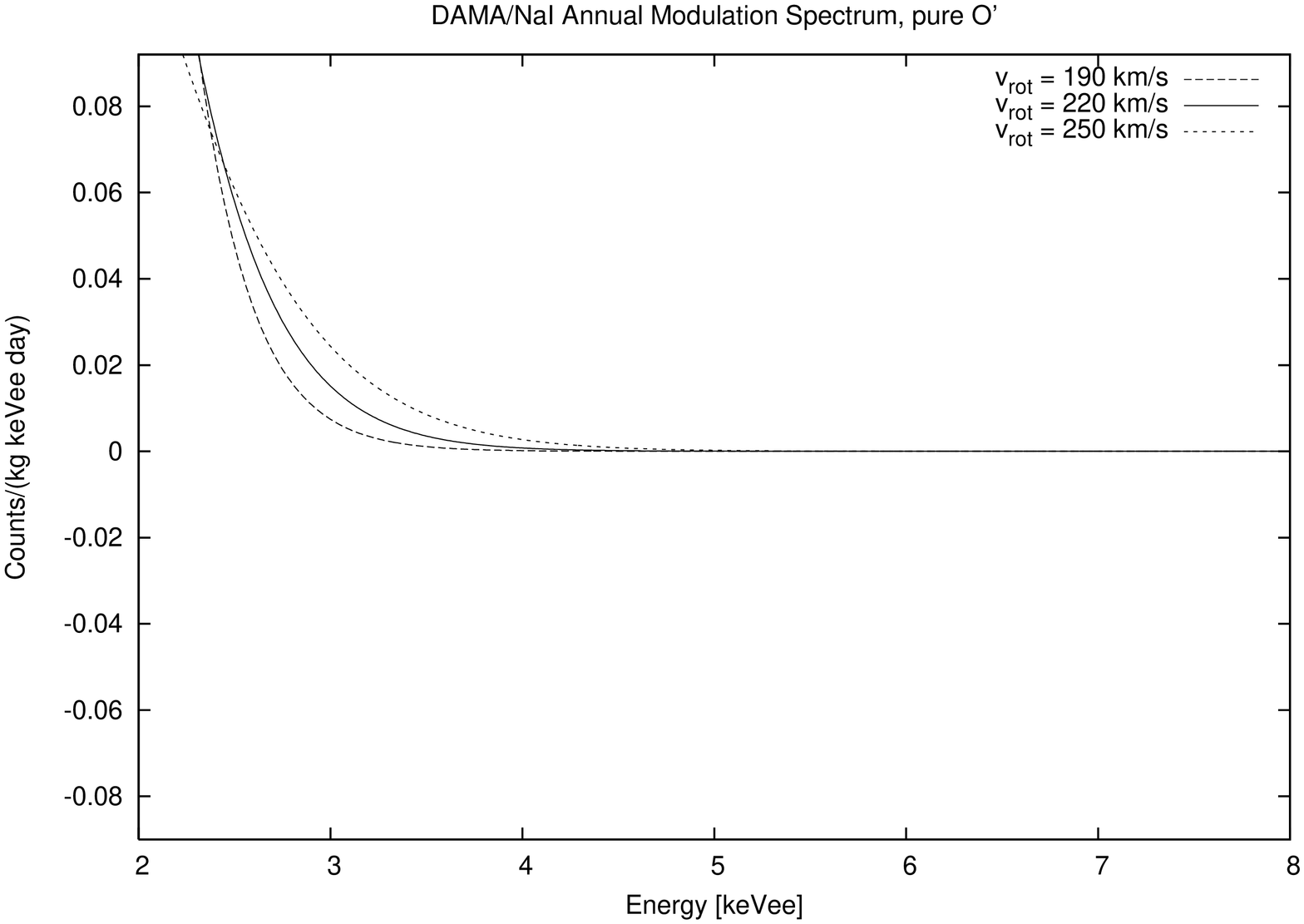,angle=270,width=12.6cm}}
\vskip 0.4cm
\noindent
Figure 1a: DAMA/NaI annual modulation energy spectrum
(as defined in text)
with $\xi_{Si'} = \xi_{Fe'} =
0$ and $|\epsilon |
\sqrt{\xi_{O'}}$ fixed so that $R^1 = 0.019$ cpd/kg/keVee in the 2-6 keVee region.
\vskip 0.5cm
\centerline{\epsfig{file=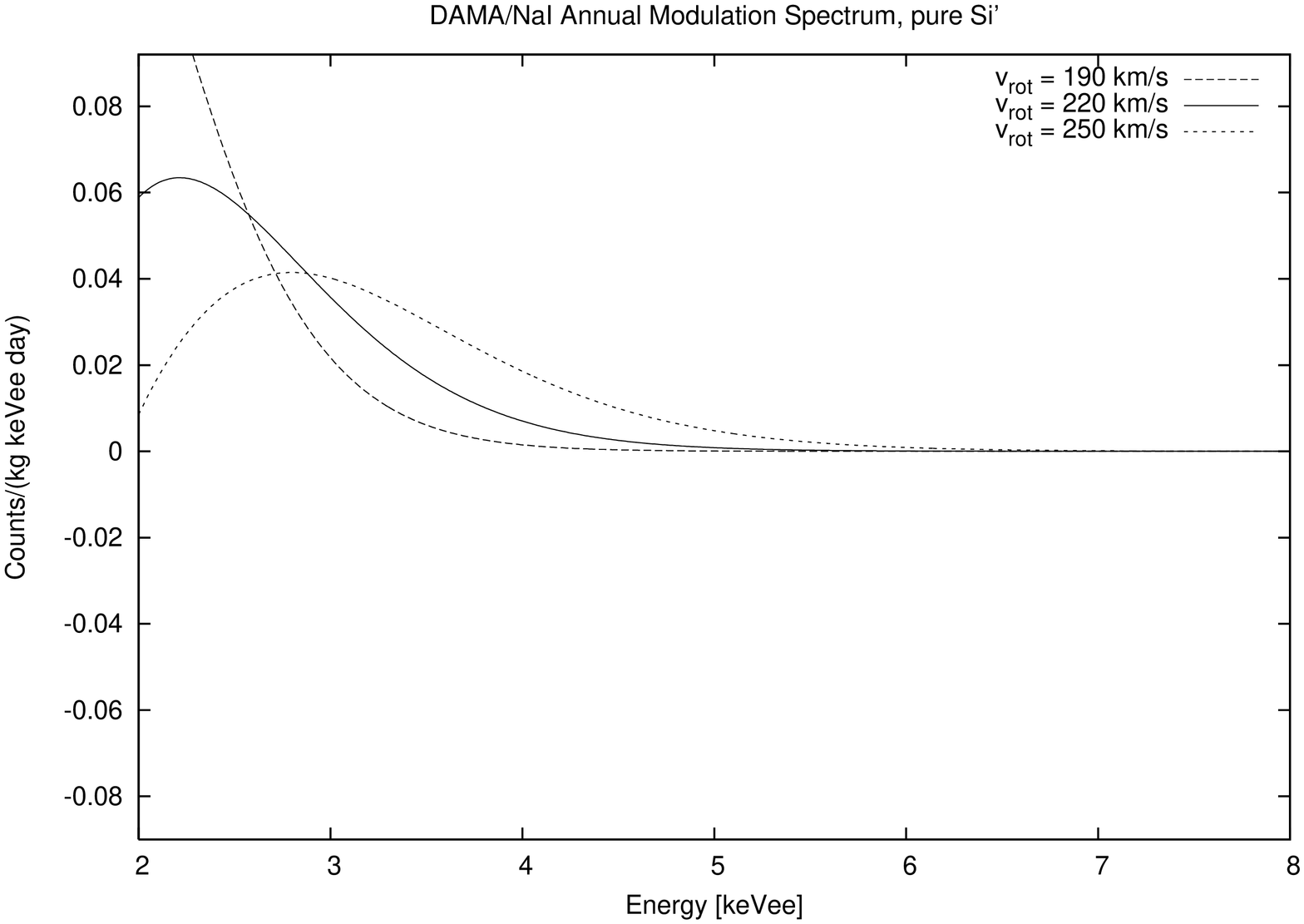,angle=270,width=12.6cm}}
\vskip 0.5cm
\noindent
Figure 1b: DAMA/NaI annual modulation energy spectrum, with $\xi_{O'} = \xi_{Fe'} =
0$ and $|\epsilon |
\sqrt{\xi_{Si'}}$ fixed so that $R^1 = 0.019$ cpd/kg/keVee in the 2-6 keVee region.
\vskip 0.5cm
\centerline{\epsfig{file=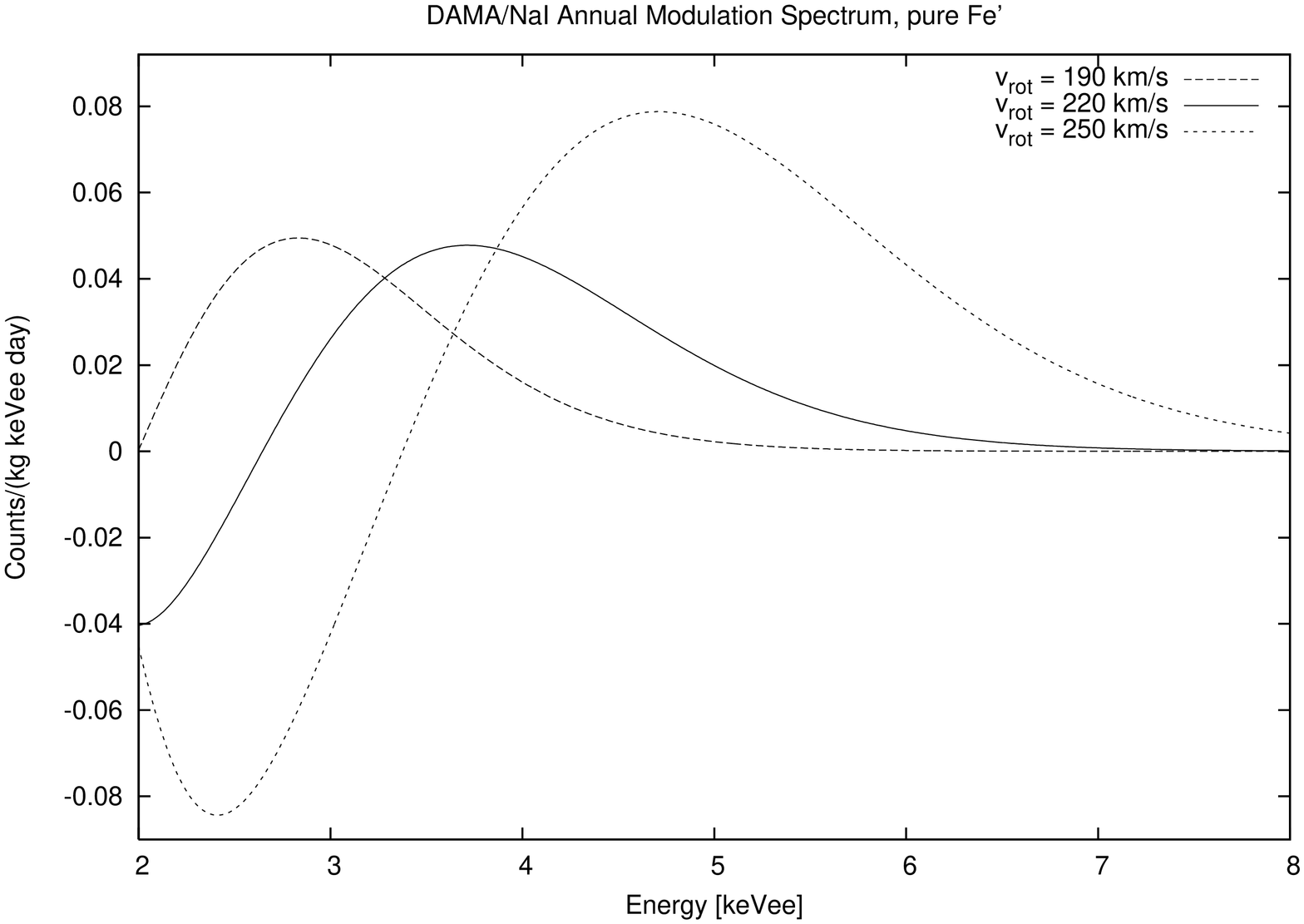,angle=270,width=12.6cm}}
\vskip 0.5cm
\noindent
Figure 1c: DAMA/NaI annual modulation energy spectrum, with $\xi_{O'} = \xi_{Si'} =
0$ and $|\epsilon |
\sqrt{\xi_{Fe'}}$ fixed so that $R^1 = 0.019$ cpd/kg/keVee in the 2-6 keVee region.
\vskip 1cm

As the figures illustrate, the spectrum for the pure $O'$ case is very steep, with 
negligible annual modulation in the (4-6) keVee region. $Si'$ and 
especially the $Fe'$ case, on the other hand,
are much flatter. Of course, this behaviour is very easy to understand: 
the heavier elements, impacting
with the Earth with typical velocity, $\sim v_{rot}$, can transfer 
more momentum to the target 
nuclei, and thus can give a significant signal at larger recoil energies. 
Importantly, there is a negligible region of parameter space which gives
a significant
annual modulation above 6 keVee --  consistent with the DAMA/NaI 
experiment which only observed an annual modulation below 6
keVee.\cite{dama}

Information about the annual modulation energy spectrum 
can be obtained from the published measurements\cite{dama}:
\begin{eqnarray}
R^1[(2-4){\rm keVee}] = 0.0233 \pm 0.0047\ {\rm cpd/kg/keVee} \nonumber \\
R^1[(2-5){\rm keVee}] = 0.0210 \pm 0.0038\ {\rm cpd/kg/keVee} \nonumber \\
R^1[(2-6){\rm keVee}] = 0.0192 \pm 0.0031\ {\rm cpd/kg/keVee} 
\end{eqnarray}
In particular, we can infer from the above that the annual modulation is
likely to be non-negligible in the (4-6) keVee region: 
\begin{eqnarray}
R^1[(4-6){\rm keVee}] = 0.015 \pm 0.005\ {\rm cpd/kg/keVee}
\label{kej}
\end{eqnarray} 
Clearly, this suggests that $Si'$ and/or $Fe'$ are non-negligible component(s).
This can be quantified, by considering the (approximate) $2\sigma$ range:
$0.005 < R^1[(4-6){\rm keVee}] < 0.025$, with $A[(2-6){\rm keVee}] = 0.019$. This
is equivalent to:
\begin{eqnarray}
0.26  < R^1[(4-6){\rm keVee}]/R^1[(2-6){\rm keVee}] < 1.32 
\label{xd}
\end{eqnarray}
In {\bf figure 2}, we plot the allowed region
of parameter space consistent with this constraint.
We vary $\xi_{Si'}$, for 3 fixed values for $\xi_{Fe'}$.
\vskip 0.5cm
\centerline{\epsfig{file=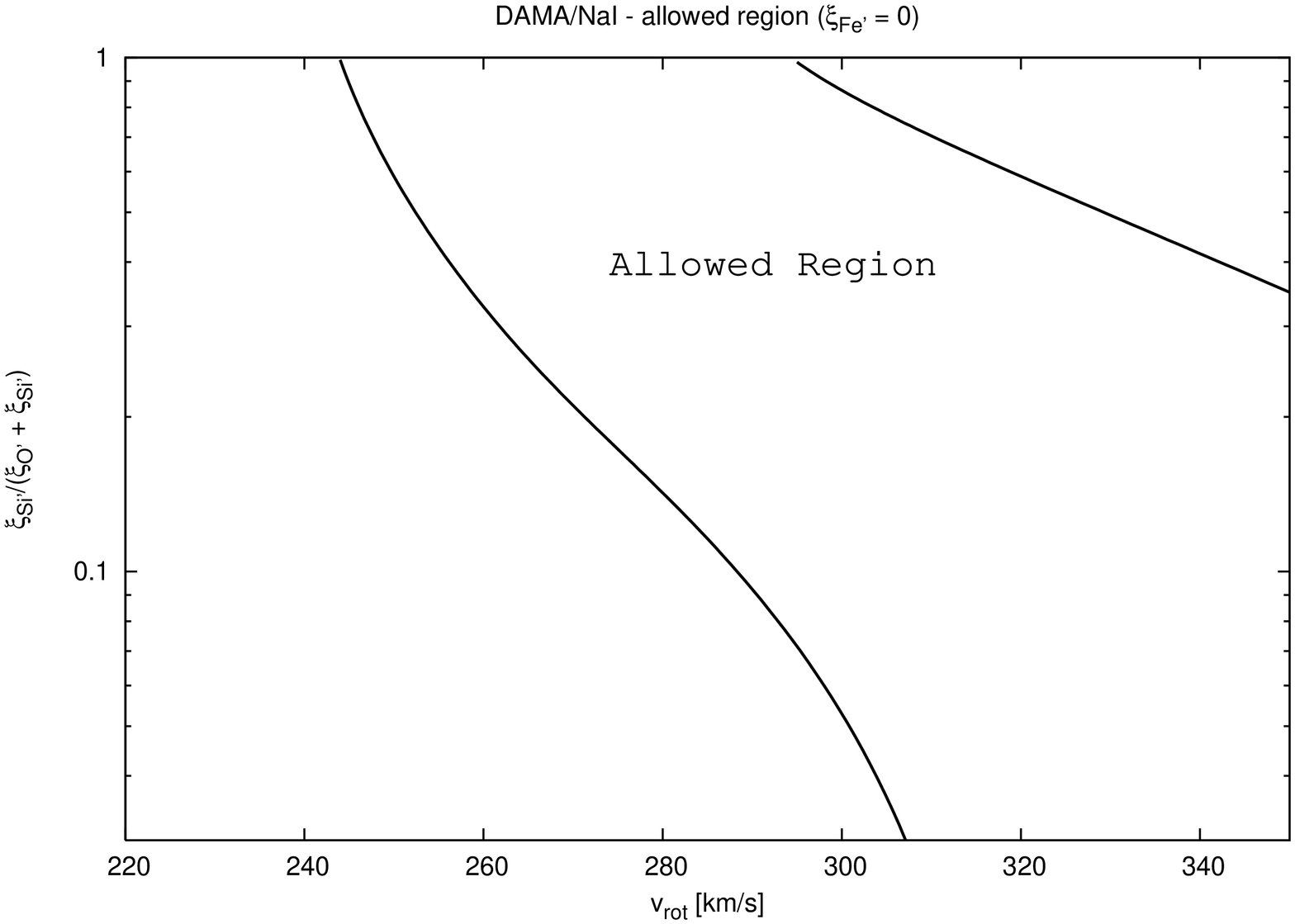,angle=270,width=12.6cm}}
\vskip 0.5cm
\noindent
Figure 2a: Region of parameter space ($\sim 2 \sigma$ allowed region) consistent with 
the DAMA/NaI annual modulation energy spectrum constraint,
$0.26  < R^1[(4-6){\rm keVee}]/R^1[(2-6){\rm keVee}] < 1.32$. This figure assumes
$\xi_{Fe'} = 0$. 
 
\vskip 0.5cm
\centerline{\epsfig{file=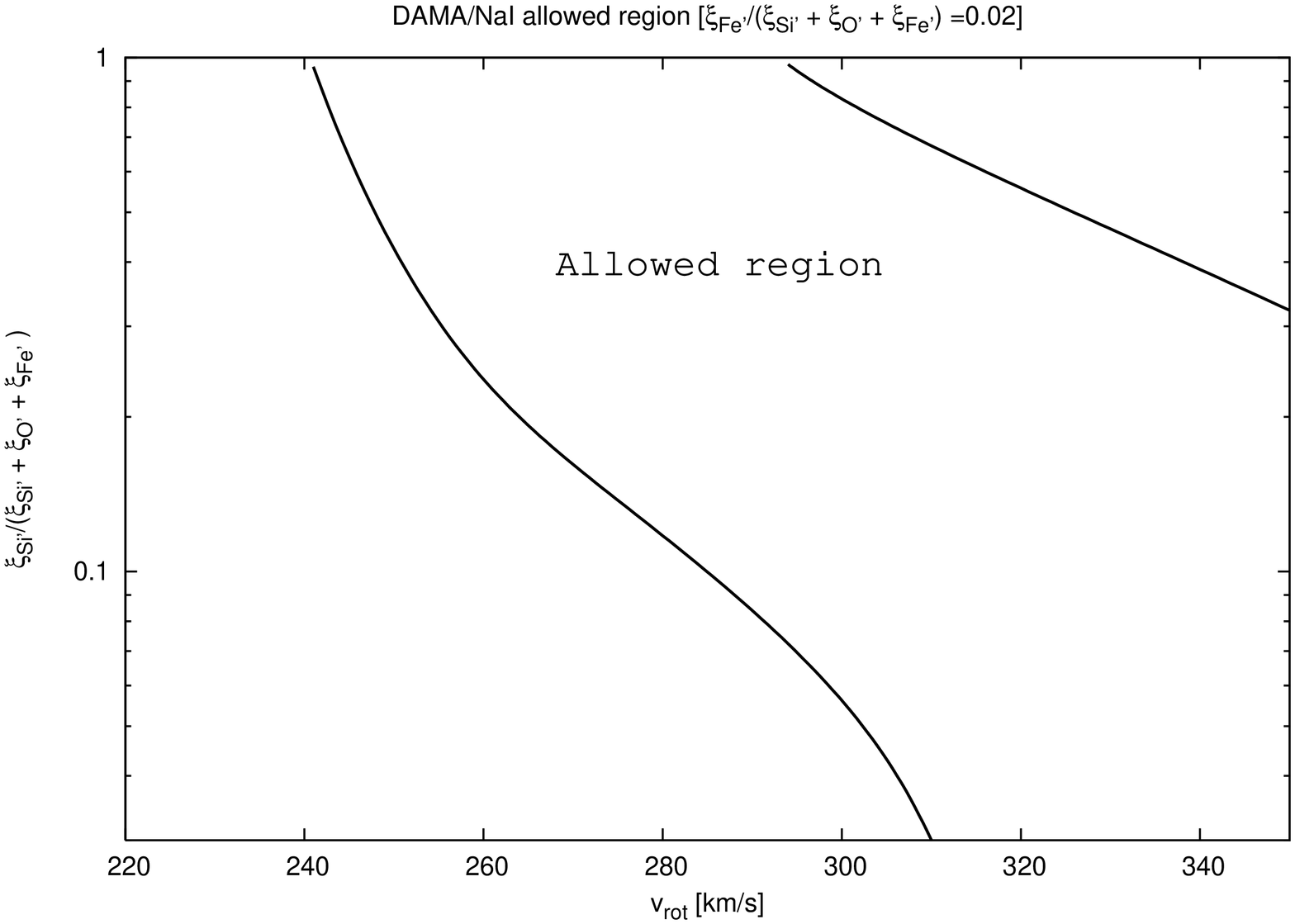,angle=270,width=12.6cm}}
\vskip 0.5cm
\noindent
Figure 2b: Same as figure 2a, except, $\xi_{Fe'}/(\xi_{O'} + \xi_{Si'} + \xi_{Fe'}) =
0.02$. 
\vskip 0.5cm
\centerline{\epsfig{file=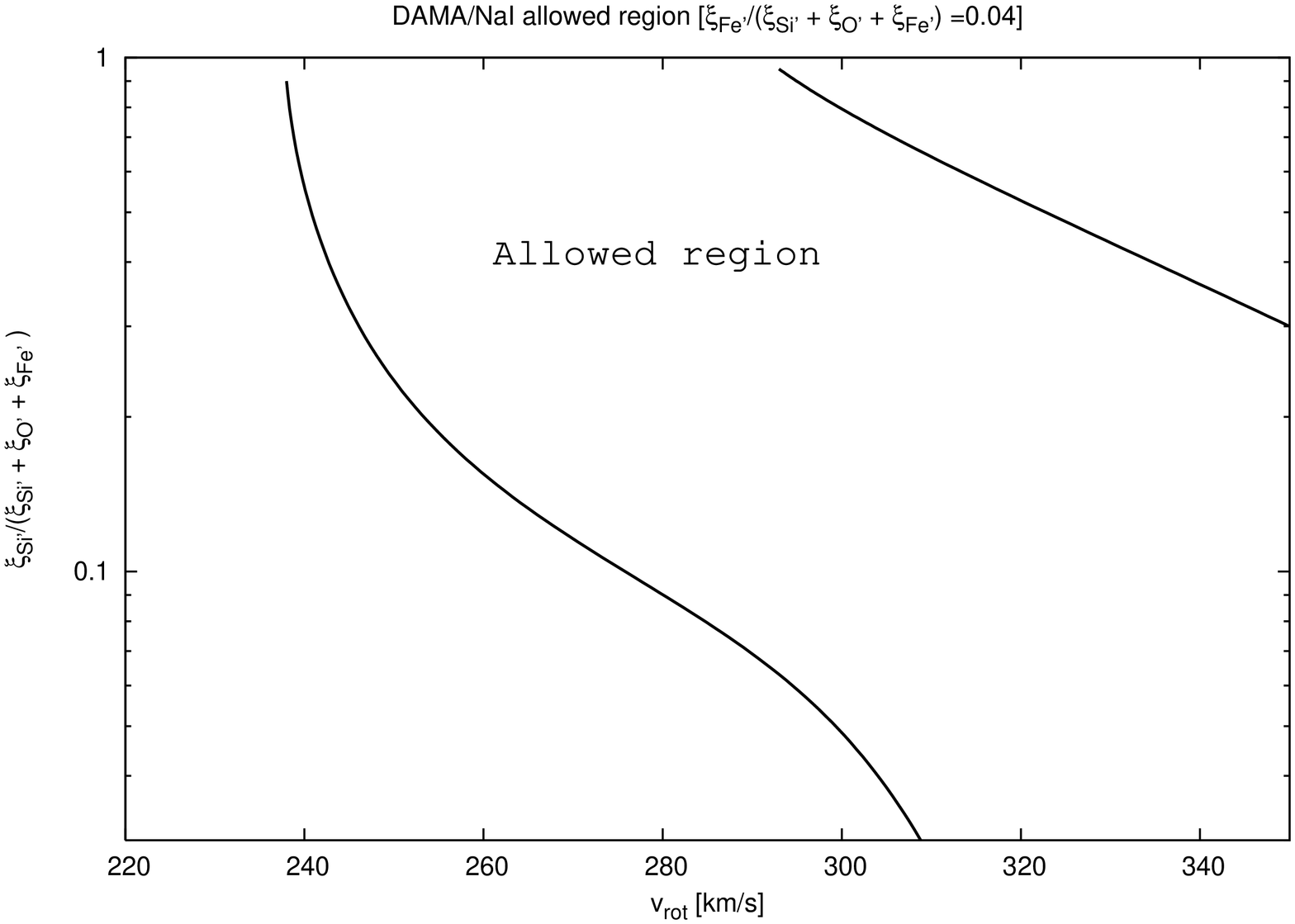,angle=270,width=12.6cm}}
\vskip 0.5cm
\noindent
Figure 2c: Same as figure 2a, except, $\xi_{Fe'}/(\xi_{O'} + \xi_{Si'} + \xi_{Fe'}) =
0.04$. 

\vskip 1.0cm


\section{The CDMS/Ge experiment}

We now turn to the CDMS/Ge experiment\cite{cdms}.
This experiment has a threshold of $10$ keV with a germanium target.
Unlike the DAMA/NaI experiment, the CDMS/Ge experiment is not sensitive
to the annual modulation effect,
but aims to measure the absolute interaction rate (which we approximate
by fixing $v_E = v_{\odot}$):
\begin{eqnarray}
{dR \over dE_R}|_{v_E = v_{\odot}} &=& 
\sum_{A'} N_T n_{A'} \int {d\sigma \over dE_R} {f_{A'}(v,v_E=v_{\odot}) \over k} |v|
d^3v \nonumber \\
&=& \sum_{A'} N_T n_{A'}
{\lambda \over E_R^2 } \int^{\infty}_{|v| > v_{min}
(E_R)} {f_{A'}(v,v_E=v_{\odot}) \over k|v|} d^3 v 
\label{551}
\end{eqnarray}
The CDMS event rate is the product of the interaction rate and
over-all detection efficiency.
With 52.6 kg-days of raw exposure, they obtained no events
passing their detection criteria. 
Using their 
published detection efficiency (figure 3 of
ref.\cite{cdms}),
We can predict the expected number of events, 
fixing $|\epsilon| \sqrt{\xi}$ using
the positive DAMA/NaI annual modulation signal. 
Of course, the prediction depends
on the chemical composition of the halo. In {\bf figure 3}, 
we give the three illustrative cases of
a $He'$ dominated halo with mirror metal component consisting of:
a) pure $O'$ [figure 3a], b) pure $Si'$ [figure 3b] and c) pure $Fe'$ [figure 3c].
\vskip 0.5cm
\centerline{\epsfig{file=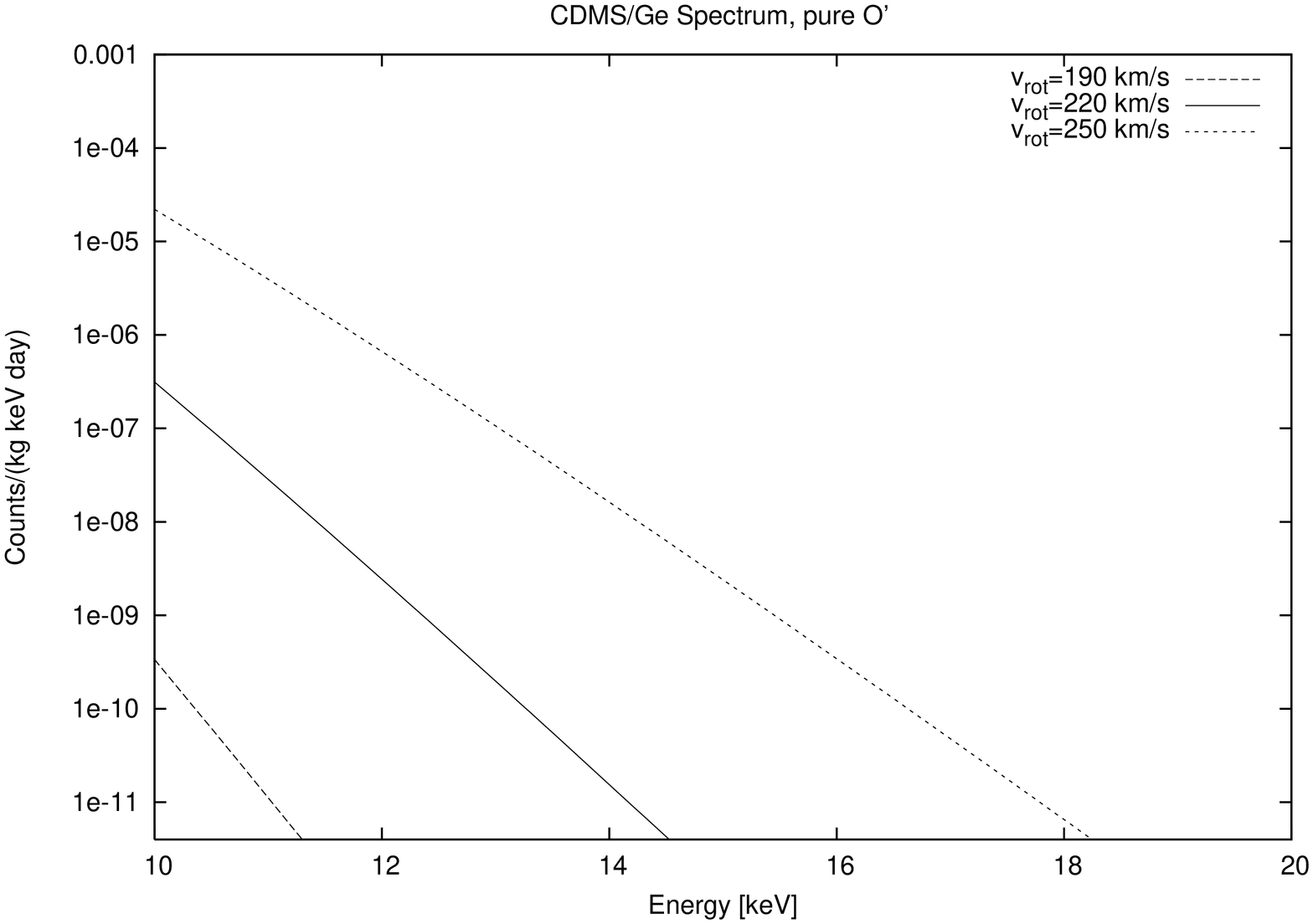,angle=270,width=12.4cm}}
\vskip 0.3cm
\noindent
Figure 3a: Predicted CDMS/Ge energy spectrum: $\epsilon (E_R) \times
\frac{dR}{dE_R}|_{v_E = v_{\odot}}$ (where $\epsilon (E_R)$ 
is the published detection
efficiency) for $\xi_{Si'} = \xi_{Fe'} = 0$ 
and $|\epsilon |\sqrt{\xi_{O'}}$ fixed by the positive
DAMA/NaI annual modulation signal, $R^1 [(2-6){\rm keVee}] = 0.019$ cpd/kg/keVee.
\vskip 0.2cm
\centerline{\epsfig{file=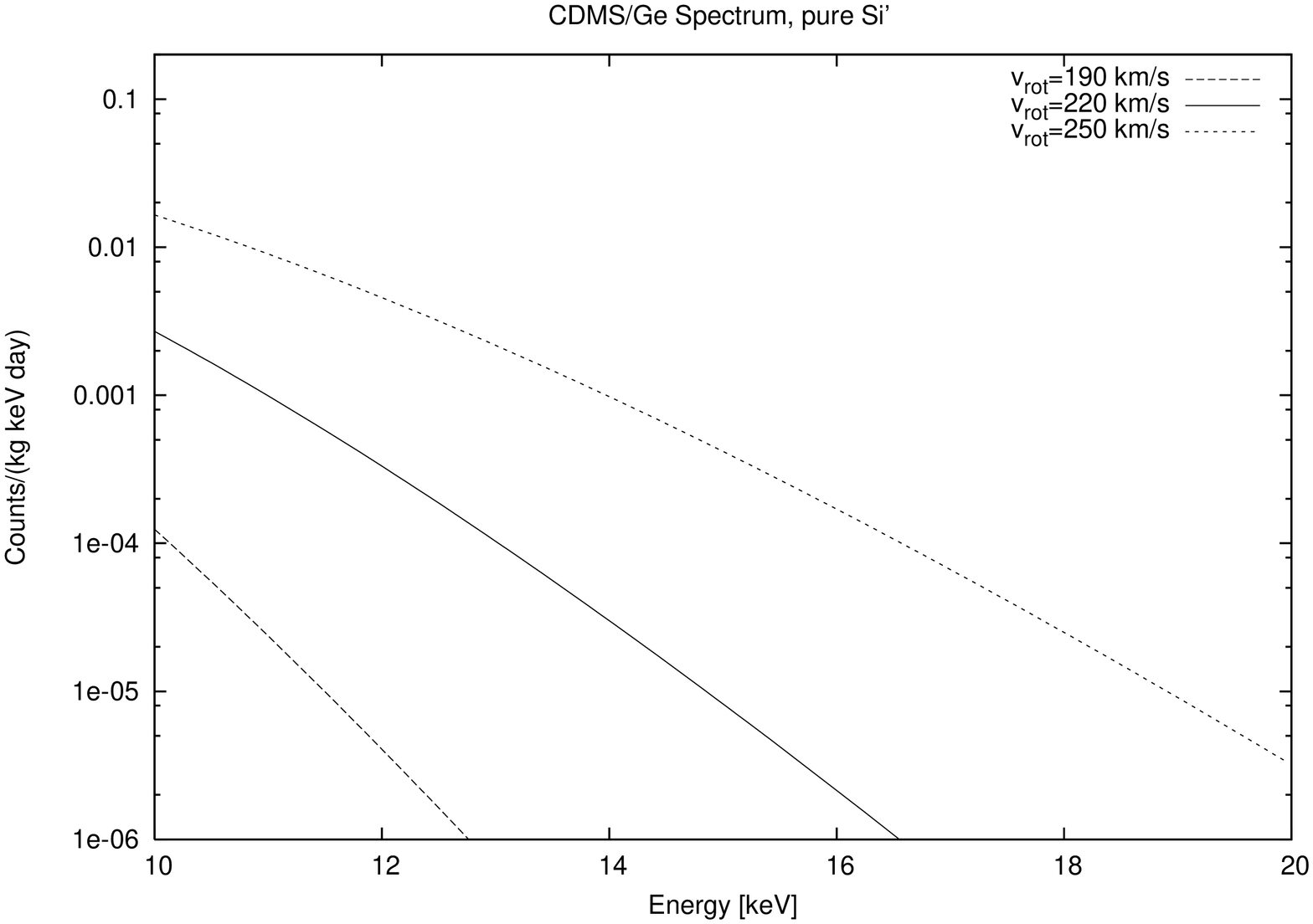,angle=270,width=12.4cm}}
\vskip 0.2cm
\noindent
Figure 3b: Same as figure 3a, except that $\xi_{O'} = \xi_{Fe'} = 0$
and $|\epsilon|\sqrt{\xi_{Si'}}$ is fixed by the positive
DAMA/NaI annual modulation signal, $R^1 [(2-6){\rm keVee}] = 0.019$ cpd/kg/keVee.
\vskip 0.2cm
\centerline{\epsfig{file=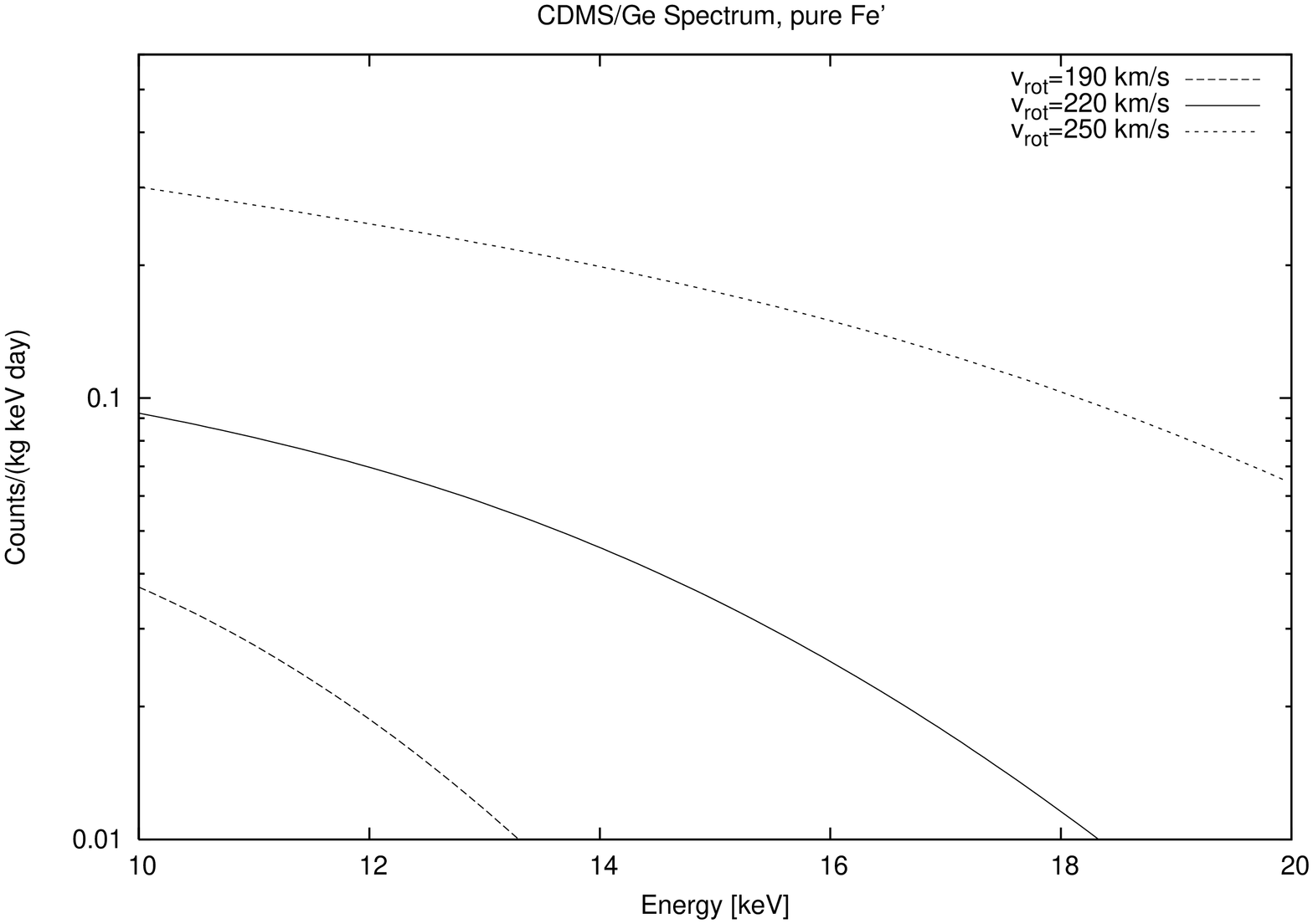,angle=270,width=12.4cm}}
\vskip 0.2cm
\noindent
Figure 3c: Same as figure 3a, except that $\xi_{O'} = \xi_{Si'} = 0$
and $|\epsilon|\sqrt{\xi_{Fe'}}$ is fixed by the positive
DAMA/NaI annual modulation signal, $R^1 [(2-6){\rm keVee}] = 0.019$ cpd/kg/keVee.
\vskip 1.0cm

As the figures show, the CDMS/Ge experiment is completely insensitive
to $O'$, but does have some significant sensitivity to the $Si'$ 
and $Fe'$ components.
Of course, the reason for this is 
clear: the heavier elements can transfer 
more momentum to the target nuclei and can therefore give more events.

Recall, a heavy $Si', Fe'$ component is expected from the
non-negligible annual modulation
in the $4-6$ keVee region observed in the DAMA/NaI experiment.
Therefore, the current null results from the
CDMS/Ge experiment does significantly constrain the
mirror matter interpretation of the DAMA experiment.
The null result of CDMS/Ge suggests a limit of $N < 3$ (at 95\% C.L.) for
their 52.6 kg-day sample (or equivalently, less than 0.057 cpd/kg). 
In {\bf figure 4} we combine this CDMS/Ge limit with the DAMA/NaI constraint, 
Eq.(\ref{xd}),
to give the combined DAMA/NaI-CDMS/Ge allowed regions
in the $v_{rot}$, $\xi_{Si'}/(\xi_{Si'} + \xi_{O'} + \xi_{Fe'})$ plane.
Figure 4a assumes $\xi_{Fe'}/(\xi_{Si'} + \xi_{O'} + \xi_{Fe'}) = 0$, 
while figure 4b assumes $\xi_{Fe'}/(\xi_{Si'} +\xi_{O'} + \xi_{Fe'}) = 0.02$, 
and figure 4c assumes $\xi_{Fe'}/(\xi_{Si'} +\xi_{O'} + \xi_{Fe'}) = 0.04$. 
There is no allowed parameter space for $\xi_{Fe'}/(\xi_{Si'}+\xi_{O'}+\xi_{Fe'}) >
0.10$.
\vskip 0.5cm
\centerline{\epsfig{file=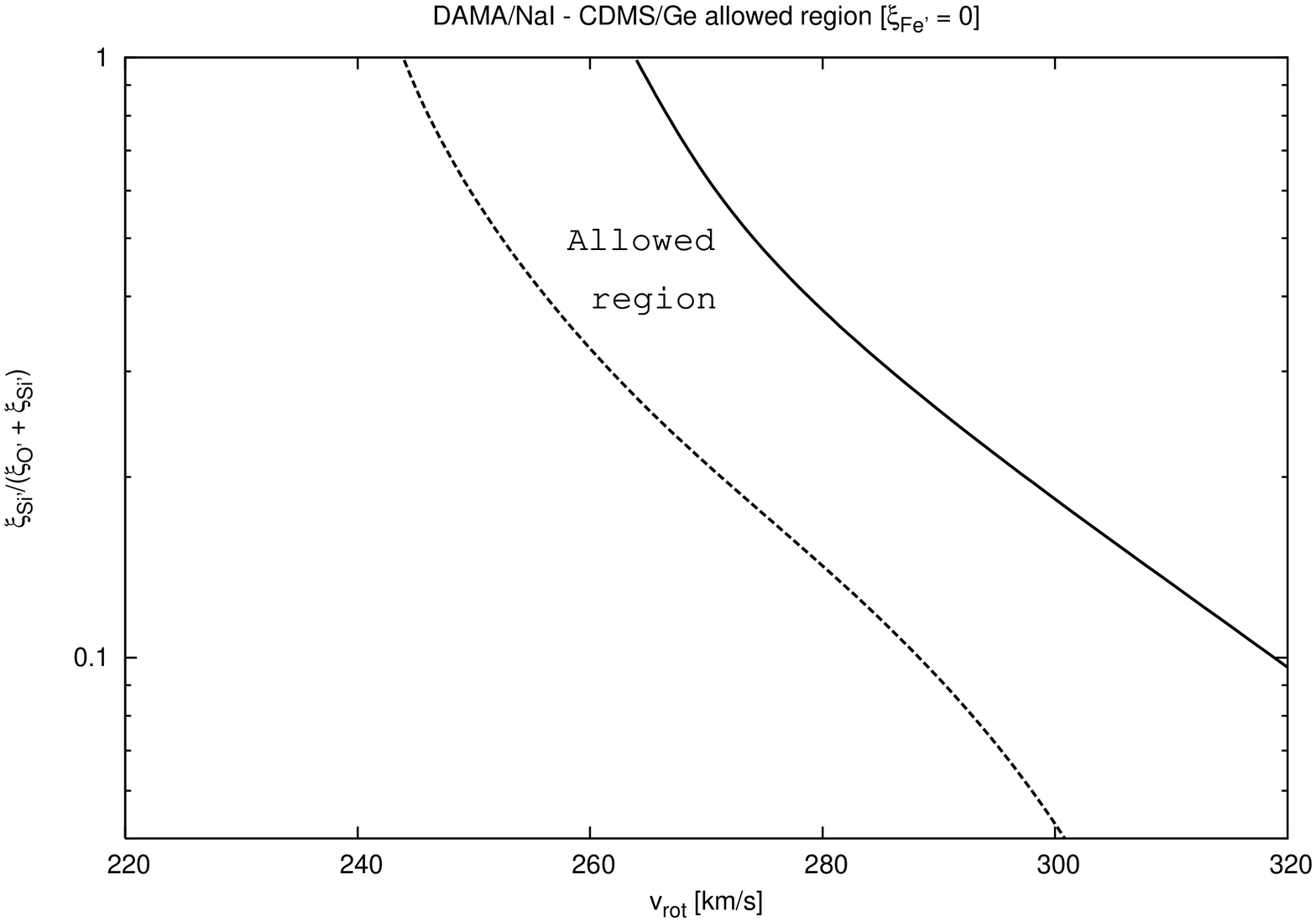,angle=270,width=12.6cm}}
\vskip 0.3cm
\noindent
Figure 4a: Region of parameter space (allowed region) 
consistent with the DAMA/NaI
annual modulation signal,
annual modulation energy spectrum constraint, Eq.(\ref{xd}) and 
also consistent with the null results of CDMS/Ge (at about 95\% C.L.).
This figure assumes $\xi_{Fe'} = 0$.
The region to the right of the dashed curve corresponds to 
the DAMA/NaI lower limit,
$R^1[(4-6)keVee]/R^1[(2-6)keVee] > 0.26$, already given in figure 2,
while the region to the left of the solid curve 
is the CDMS/Ge constraint (predicted event 
rate $< 0.057$
cpd/kg, as discussed in text).
\vskip 0.3cm
\centerline{\epsfig{file=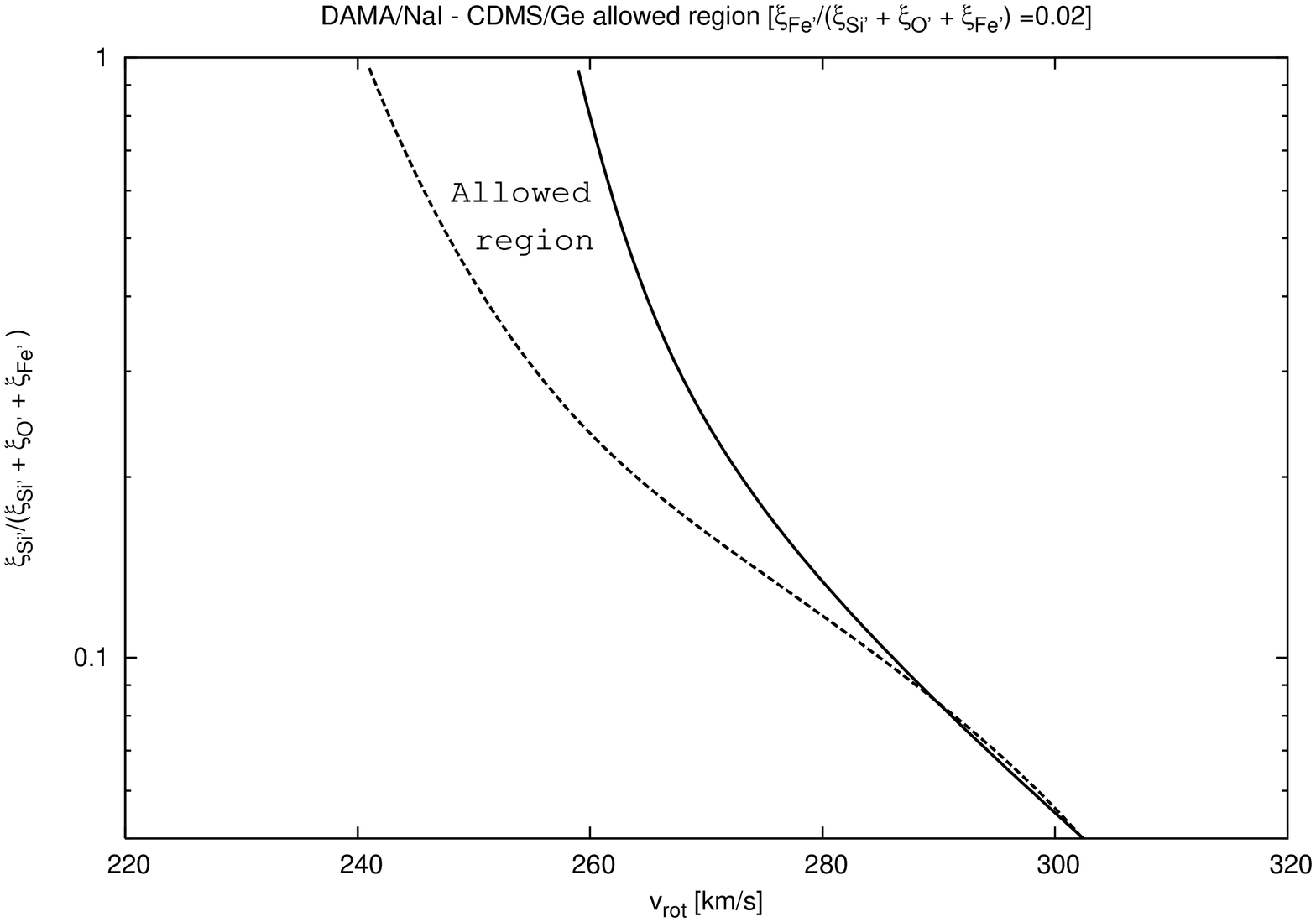,angle=270,width=12.6cm}}
\vskip 0.3cm
\noindent
Figure 4b: Same as figure 4a, except with
$\xi_{Fe'}/(\xi_{O'} + \xi_{Si'} + \xi_{Fe'}) = 0.02$.
\vskip 0.5cm
\centerline{\epsfig{file=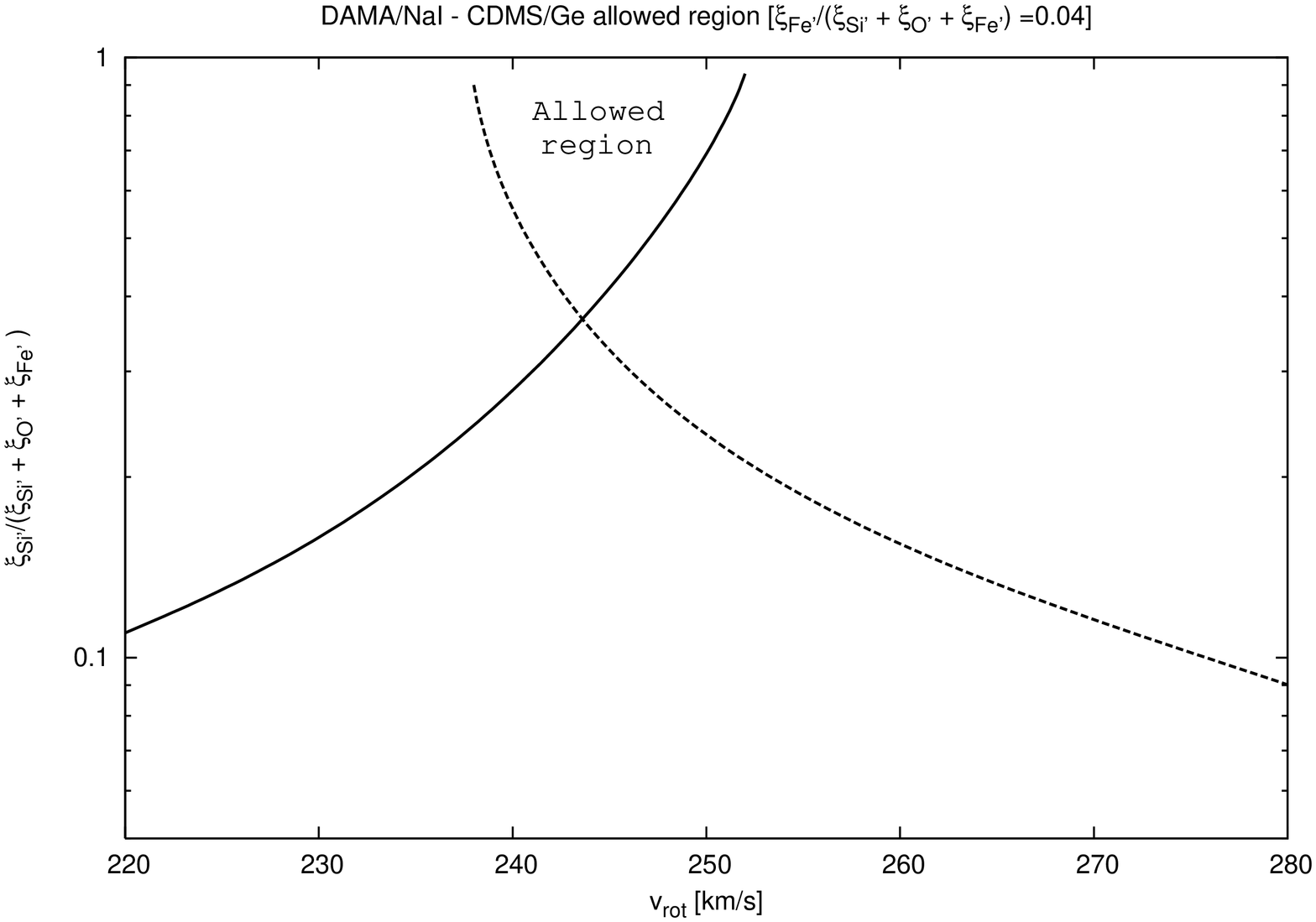,angle=270,width=12.6cm}}
\vskip 0.5cm
\noindent
Figure 4c: Same as figure 4a, except with
$\xi_{Fe'}/(\xi_{O'} + \xi_{Si'} + \xi_{Fe'}) = 0.04$.
\vskip 1.5cm
Note that the allowed parameter space is fairly well defined. In fact,
we would expect a positive signal from the CDMS/Ge experiment in the very near
future, which we predict to be at recoil energies 
very close to threshold (certainly $< 15$ keV).

Very recently, the CDMS collaboration have presented new
data\cite{cdms2} consisting of a raw exposure of about 93 kg-days for CDMS/Ge. 
Interestingly, they did obtain 1 event with a recoil energy of $10.5$ keV 
passing their detection
criteria. Furthermore, the background in the near threshold region
($E_R < 15$ keV)
is expected to be much less than 1 event -- given that the estimated background for the
$10 \ {\rm keV} < E_R < 100 \ {\rm keV}$ region is just 0.4 events\cite{cdms2}.  Of course, one
shouldn't take one event too seriously so we must wait for
confirmation or non-confirmation in the near future.
They also presented results for CDMS/Si which is also
potentially sensitive to mirror matter-type dark matter.
However, they have not published their over-all detection efficiency 
for CDMS/Si, so a quantitative analysis is not possible.
However, assuming that the CDMS/Si experiment has the same over-all detection efficiency 
as the CDMS/Ge experiment (as given in figure 3 of ref.\cite{cdms}), then
we predict between 1-5 events for the exposure time of 74.5 live days for the allowed
regions in figure 4.

\section{Conclusion}

In conclusion, we have re-analysed the mirror matter interpretation of the positive
dark matter signal obtained in the DAMA/NaI experiment, taking into account the
annual modulation spectrum constraint, Eq.(\ref{xd}). We have combined this with
the null results from the CDMS/Ge experiment, to yield fairly well defined
allowed regions of parameter space (figure 4). This favoured parameter space will be
probed in the near future by the currently running DAMA/LIBRA and CDMS experiments.
In particular, this interpretation of the DAMA/NaI experiment suggests 
that the CDMS/Ge (and CDMS/Si) experiment(s) should see a positive signal around the 
recoil energy threshold $E_R < 15 \ {\rm keV}$
in the near future.

\vskip 1cm
\noindent
{\bf Acknowledgements}
\vskip 0.4cm
\noindent
This work was supported by the Australian Research Council.


\begin{thebibliography}{999}

\bibitem{flv}
R. Foot, H. Lew and R. R. Volkas, Phys. Lett. B272, 67 (1991).
The idea was earlier discussed, prior to the
advent of the standard model, in:
T. D. Lee and C. N. Yang, Phys. Rev. 104, 256 (1956);
I. Kobzarev, L. Okun and I. Pomeranchuk, Sov. J. Nucl. Phys. 3, 837
(1966).

\bibitem{review}
For an up-to-date review, see
R. Foot, Int. J. Mod. Phys. D13, 2161 (2004)
[astro-ph/0407623].

\bibitem{comelli}
Z. Berezhiani, D. Comelli and F. L. Villante, 
Phys. Lett. B503, 362 (2001) [hep-ph/0008105].

\bibitem{lss}
A. Yu. Ignatiev and R. R. Volkas, Phys. Rev. D68, 023518 (2003)
[hep-ph/0304260].

\bibitem{other}
Z. Berezhiani, P. Ciarcelluti, D. Comelli and F. L. Villante,
Int. J. Mod. Phys. D14, 107 (2005) [astro-ph/0312605];
P. Ciarcelluti, Int. J. Mod. Phys. D14, 187 (2005) [astro-ph/0409630];
Int. J. Mod. Phys. D14, 223 (2005) [astro-ph/0409633].

\bibitem{new1}
L. Bento and Z. Berezhiani, Phys. Rev. Lett. 87, 231304 (2001)
[hep-ph/0107281]; hep-ph/0111116.

\bibitem{new2}
R. Foot and R. R. Volkas, Phys. Rev. D68, 021304 (2003)
[hep-ph/0304261]; Phys. Rev. D69, 123510 (2004) [hep-ph/0402267].

\bibitem{macho} 
C. Alcock {\it et al}., 
Astrophys. J. 542, 281 (2000) [astro-ph/0001272];
R. Uglesich {\it et al}.,  Astrophys. J. 612, 877 (2004)
[astro-ph/0403248]; S. Calchi Novati {\it et al}., astro-ph/0504188.

\bibitem{sph}
R. Foot and R. R. Volkas,
Phys. Rev. D70, 123508 (2004) [astro-ph/0407522].

\bibitem{fh}
R. Foot and X-G. He, Phys. Lett. B267, 509 (1991).

\bibitem{flv2}
R. Foot, H. Lew and R. R. Volkas, Mod. Phys. Lett. A7, 2567 (1992);
R. Foot, Mod. Phys. Lett. A9, 169 (1994) [hep-ph/9402241];
R. Foot and R. R. Volkas, Phys. Rev. D52, 6595 (1995) [hep-ph/9505359].

\bibitem{holdom}
B. Holdom, Phys. Lett. B166, 196 (1986).

\bibitem{sasha}
R. Foot, A. Yu. Ignatiev and R. R. Volkas,
Phys. Lett. B503, 355 (2001) [astro-ph/0011156].

\bibitem{freview}
R. Foot, Int. J. Mod. Phys. A19 3807 (2004) [astro-ph/0309330].

\bibitem{f1}
R. Foot, Phys. Rev. D69, 036001 (2004) [hep-ph/0308254].

\bibitem{f2}
R. Foot, astro-ph/0403043.

\bibitem{f3}
R. Foot, Mod. Phys. Lett. A19, 1841 (2004) [astro-ph/0405362].

\bibitem{dama}
R. Bernabei et al. (DAMA Collaboration), 
Phys. Lett. B480, 23 (2000);
Riv. Nuovo Cimento. 26, 1 (2003) [astro-ph/0307403]; Int. J. Mod.
Phys. D13, 2127 (2004) and references there-in.

\bibitem{cdms}
D. S. Akerib {\it et al}. (CDMS Collaboration),
Phys. Rev. Lett. 93, 211301 (2004)
[astro-ph/0405033].

\bibitem{helm}
R. H, Helm, Phys. Rev. 104, 1466 (1956). 

\bibitem{smith}
J. D. Lewin and P. F. Smith, Astropart. Phys. 6, 87 (1996).

\bibitem{koch}
C. S. Kochanek, Astrophys. J. 457, 228 (1996).


\bibitem{sig}
A. K. Drukier, K. Freese and D. N. Spergel, Phys. Rev. D33, 3495 (1986);
K. Freese, J. A. Frieman 
and A. Gould,  Phys. Rev. D37, 3388 (1988).

\bibitem{recent}
Z. Berezhiani, S. Cassisi, P. Ciarcelluti and A. Pietrinferni, astro-ph/0507153.

\bibitem{recent2}
The experimental energy resolution is taken into account by 
convolving the rate with
a gaussian, with $\sigma$ obtained from figure 3 of 
R. Bernabei {\it et al}., Eur. Phys. J. C 18, 283 (2000).

\bibitem{saibal}
R. Foot and S. Mitra, Astropart. Phys. 19, 739 (2003)
[astro-ph/0211067].

\bibitem{zurab}
R. Foot and Z. K. Silagadze, Int. J. Mod. Phys. D14, 143 (2005)
[astro-ph/0404515].

\bibitem{shelly}
S. L. Glashow, Phys. Lett. B167, 35 (1986); R. Foot and S. N. Gninenko,
Phys. Lett. B480, 171 (2000) [hep-ph/0003278]. Although the current
best limit on $\epsilon$ from orthopositronium experiments 
is\cite{freview} $\epsilon < 5\times 10^{-7}$, forthcoming experiments [A.
Badertscher {\it et al}., Int. J. Mod. Phys. A19, 3833 (2004)]
may probe the interesting region $\epsilon \sim 10^{-9}$.

\bibitem{shelly2}
E. D. Carlson and S. L. Glashow, Phys. Lett. B193, 168 (1987).

\bibitem{cdms2}
D. S. Akerib {\it et al}. (CDMS Collaboration),
astro-ph/0509259.





\end{thebibliography}
\end{document}